\documentclass[trackchanges]{aastex701}

\usepackage{multirow}
\usepackage[table,xcdraw]{xcolor}
\usepackage{float}
\defcitealias{2022PASP..134k4501C}{CASA Team et al. 2022}
\usepackage{natbib}
\begin{document}

\title{ALMA Observations of DEM L241/LMC P3 in the Large Magellanic Cloud:
Evidence for the Formation of Cool Molecular Jets Driven by a Microquasar 
}

\author[]{Yasuo Fukui}
\affiliation{Department of Physics, Nagoya University, Furo-cho, Chikusa-ku, Nagoya, Aichi 464-8601, Japan}
\affiliation{Faculty of Engineering, Gifu University, 1-1 Yanagido, Gifu 501-1193, Japan}
\email{fukui@a.phys.nagoya-u.ac.jp}

\correspondingauthor{Bhuvana G. R.}
\author[]{Bhuvana G. R.}
%\altaffiliation{Institute for Advanced Science}
\affiliation{Institute for Advanced Study, Gifu University, 1-1 Yanagido, Gifu 501-1193, Japan}
\email[show]{bhuvana.g.r.k4@f.gifu-u.ac.jp}  

\author[]{Hidetoshi Sano}
\affiliation{Faculty of Engineering, Gifu University, 1-1 Yanagido, Gifu 501-1193, Japan}
\email[]{sano.hidetoshi.w4@f.gifu-u.ac.jp}

\author[]{Kisetsu Tsuge} 
%\altaffiliation{Las Campanas Observatory}
\affiliation{Institute for Advanced Study, Gifu University, 1-1 Yanagido, Gifu 501-1193, Japan}
\affiliation{Faculty of Engineering, Gifu University, 1-1 Yanagido, Gifu 501-1193, Japan}
\affiliation{National Astronomical Observatory of Japan, National Institutes of Natural Sciences, 2-21-1 Osawa, Mitaka, Tokyo 181-8588, Japan}
\affiliation{Institute for Advanced Research, Nagoya University, Furo-cho, Chikusa-ku, Nagoya 464-8601, Japan}
\email{tsuge.kisetsu.i2@f.gifu-u.ac.jp}

\author[]{Rami Z. E. Alsaberi}
\affiliation{Faculty of Engineering, Gifu University, 1-1 Yanagido, Gifu 501-1193, Japan}
\affiliation{Western Sydney University, LockedBag1797, Penrith South DC, NSW 1797, Australia}
\email{ramy_z@yahoo.com}

\author[]{Rin Yamada}
%\altaffiliation{}
\affiliation{Faculty of Engineering, Gifu University, 1-1 Yanagido, Gifu 501-1193, Japan}
\affiliation{National Astronomical Observatory of Japan, National Institutes of Natural Sciences, 2-21-1 Osawa, Mitaka, Tokyo 181-8588, Japan}
\email{rin.yamada@nao.ac.jp}

\author[]{Yuya Asano}
\affiliation{Institute for Advanced Study, Gifu University, 1-1 Yanagido, Gifu 501-1193, Japan}
\email{asano.yuuya.m9@s.gifu-u.ac.jp}

\author[]{Aya Bamba}
\email[]{bamba@phys.s.u-tokyo.ac.jp}
\affiliation{Department of Physics, Graduate School of Science, The University of Tokyo, 7-3-1 Hongo, Bunkyo-ku, Tokyo 113-0033, Japan}
\affiliation{Research Center for the Early Universe, School of Science, The University of Tokyo, 7-3-1 Hongo, Bunkyo-ku, Tokyo 113-0033, Japan}
\affiliation{Trans-Scale Quantum Science Institute, The University of Tokyo, Tokyo 113-0033, Japan}

\author[]{Miroslav Filipovic}
\email[]{m.filipovic@westernsydney.edu.au}
\affiliation{Western Sydney University, LockedBag1797, Penrith South DC, NSW 1797, Australia}

\author[]{Charles Law}
\affiliation{Department of Astronomy, University of Virginia, Charlottesville, VA 22904, USA}
\affiliation{Minnesota Institute for Astrophysics, University of Minnesota, 116 Church Street SE, Minneapolis, MN 55455, USA}
\email[]{law06855@umn.edu}

\author[]{Norikazu Mizuno}
\affiliation{National Astronomical Observatory of Japan, Mitaka, Tokyo 181-8588, Japan}
\email[]{norikazu.mizuno@nao.ac.jp}

\author[]{Toshikazu Onishi}
\affiliation{Department of Physics, Graduate School of Science, Osaka Metropolitan University, 1-1
Gakuen-cho, Naka-ku, Sakai, Osaka 599-8531,Japan}
\email[]{tonishi@omu.ac.jp}

\author[]{Paul Plucinsky}
\affiliation{Center for Astrophysics, Harvard-Smithsonian, MA 02138, USA}
\email[]{pplucinsky@cfa.harvard.edu}

\author[]{Gavin Rowell}
\affiliation{School of Physical Sciences, University of Adelaide, Adelaide 5005, Australia}
\email[]{gavin.rowell@adelaide.edu.au}

\author[]{Manami Sasaki}
\affiliation{Friedrich-Alexander-Universität Erlangen-Nürnberg, Erlangen Centre for Astroparticle Physics, Nikolaus-Fiebiger-Str. 2, 91058, Erlangen, Germany}
\email[]{manami.sasaki@fau.de}

\author[]{Piyush Sharda}
\affiliation{Leiden Observatory, Leiden University, PO Box 9513, 2300 RA Leiden, The Netherlands}
\email[]{sharda@strw.leidenuniv.nl}

\author[]{Hiromasa Suzuki}
\affiliation{Faculty of Engineering, University of Miyazaki, 1-1 Gakuen Kibanadai Nishi, Miyazaki, Miyazaki 889-2192, Japan}
\email[]{suzuki@astro.miyazaki-u.ac.jp}

\author[]{Kengo Tachihara}
\affiliation{Department of Physics, Nagoya University, Furo-cho, Chikusa-ku, Nagoya, Aichi 464-8601, Japan}
\email[]{k.tachihara@a.phys.nagoya-u.ac.jp}

\author[]{Kazuki Tokuda}
\affiliation{Faculty of Education, Kagawa University, Saiwai-cho 1-1, Takamatsu, Kagawa 760-8522, Japan}
\email[]{tokuda.kazuki@kagawa-u.ac.jp}

\author[]{Yumiko Yamane}
\affiliation{Department of Physics, Nagoya University, Furo-cho, Chikusa-ku, Nagoya, Aichi 464-8601, Japan}
\email[]{yamane.y@a.phys.nagoya-u.ac.jp}
%\author[]{Yasuo Fukui}
%\affiliation{Department of Physics, Nagoya University, Furo-cho, Chikusa-ku, Nagoya 464-8601, Japan}
%\email{fakeemail6@google.com}

%\author[0000-0000-0000-0003,sname=Asia,gname=Mountain]{.......................}
%\altaffiliation{Astrosat Post-Doctoral Fellow}
%\affiliation{}
%\email{fakeemail5@google.com}

%\author[0000-0000-0000-0003,sname=Asia,gname=Mountain]{Takeru Murase}
%\altaffiliation{Astrosat Post-Doctoral Fellow}
%\affiliation{Tata Institute of Fundamental Research, Department of Astronomy}
%\email{fakeemail5@google.com}

%\author[gname=IceSheet]{}
%\affiliation{Amundsen–Scott South Pole Station}
%\email{fakeemail7@google.com}

%\collaboration{all}{The Terra Mater collaboration}

%% Use the \collaboration command to identify collaborations. This command
%% takes an optional argument that is either a number or the word "all"
%% which tells the compiler how many of the authors above the command to
%% show. For example "\collaboration[all]{(DELVE Collaboration)}" wil include
%% all the authors above this command.
%%
%% Mark off the abstract in the ``abstract'' environment. 
\begin{abstract}

We present ALMA observations of DEML 241/LMC P3, the most luminous $\gamma$-ray binary consisting of a compact object and an O star, in CO emission. We have found an one-sided jet-like CO feature of 8 pc length and 1 pc width, which accompanies another weaker CO jet candidate with slightly different orientation. The one-sided CO jet exhibits striking alignment with LMC P3, suggesting that the jet was driven by LMC P3. We have determined kinetic temperature of the CO jet to be significantly high at 33$-$60 K as compared with $\sim$15 K in the nearby non-jet CO cloud whereas no radiative heat source is found. We interpret that the high temperatures are due to shock heating of a microquasar jet driven by the $\gamma$-ray binary, where the compact object has an accretion disk fed by the O star winds. The CO jet matches existing predictions from magneto-hydrodynamical simulations, which show that CO jet can form from the interaction of the microquasar jet and an ambient ISM cloud. These results provide strong evidence that CO jets are a signature sculptured by microquasar jets, lending support for mass accretion in LMC P3 as the $\gamma$-ray origin. The results suggest a second case of CO jets potentially driven by a microquasar along with the CO jets in the microquasar candidate HESS J1023-575 recently identified in the Milky Way. Further, our results suggest the use of sub-mm observations for identifying microquasars, opening a new possible window for their discovery and study.
\end{abstract}

%% Keywords should appear after the \end{abstract} command. 
%% The AAS Journals now uses Unified Astronomy Thesaurus (UAT) concepts:
%% https://astrothesaurus.org
%% You will be asked to selected these concepts during the submission process
%% but this old "keyword" functionality is maintained in case authors want
%% to include these concepts in their preprints.
%%
%% You can use the \uat command to link your UAT concepts back its source.

%% From the front matter, we move on to the body of the paper.
%% Sections are demarcated by \section and \subsection, respectively.
%% Observe the use of the LaTeX \label
%% command after the \subsection to give a symbolic KEY to the
%% subsection for cross-referencing in a \ref command.
%% You can use LaTeX's \ref and \label commands to keep track of
%% cross-references to sections, equations, tables, and figures.
%% That way, if you change the order of any elements, LaTeX will
%% automatically renumber them.

\section{Introduction} 

\begin{figure}
\centering
\includegraphics[width=0.98\textwidth]{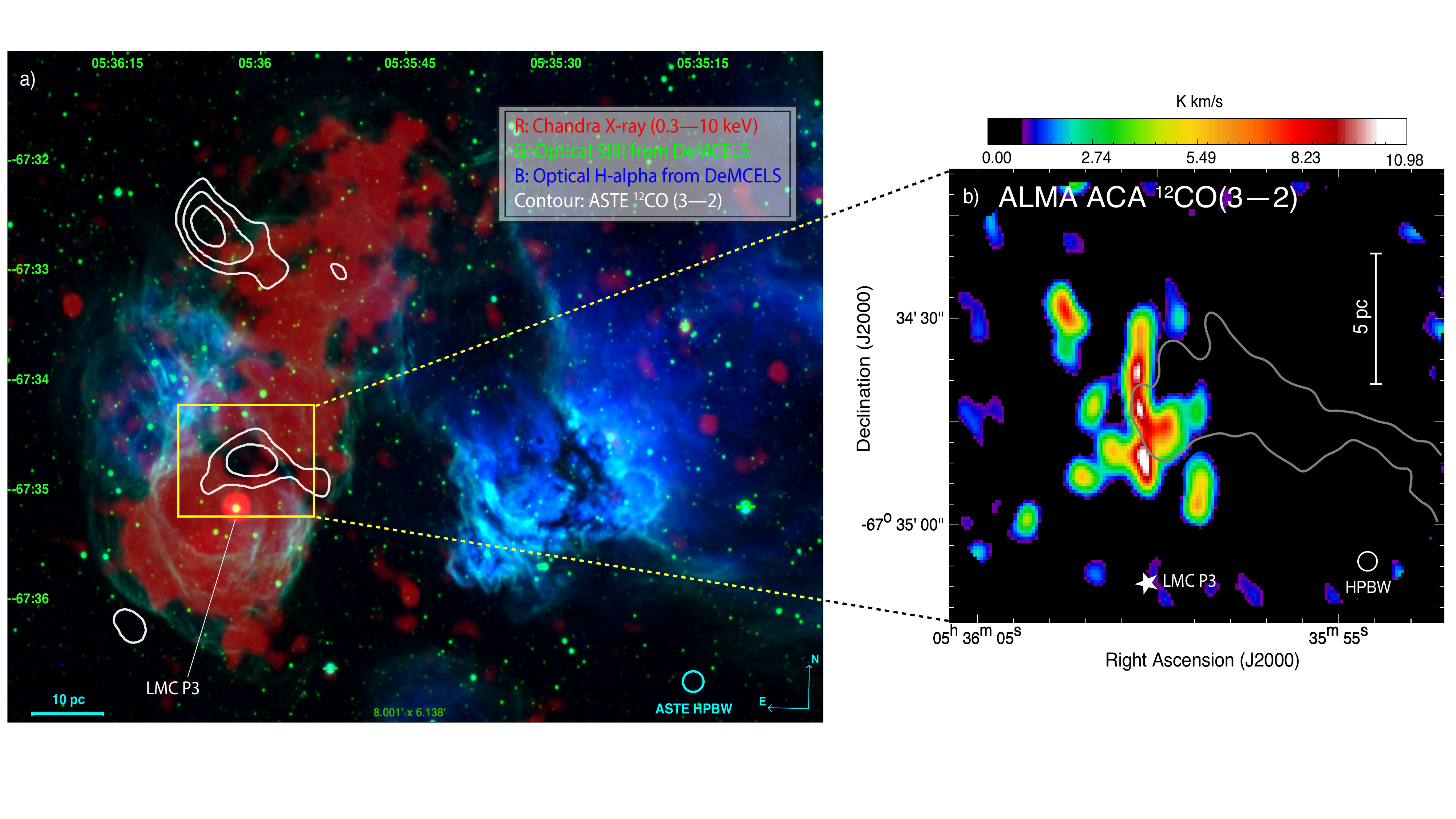}
\caption{(a) Multiwavelength overview of the SNR DEM L241 region. Red shows Chandra X-ray emission in 0.3$-$10 keV,
green shows [S {\sc ii}] optical emission from DeMCELS, and blue shows H$\alpha$ emission from DeMCELS. Overlaid contours represent ASTE $^{12}$CO($J$=3--2) emission. The yellow boxes indicate the ALMA ACA observation mosaic region. (b) ALMA $^{12}$CO($J$=3--2)
emission integrated over the velocity range 285.6$-$288.0 km s$^{-1}$ that shows jet clouds. The contour shows the ALMA $^{12}$CO($J$=3--2) emission integrated over the velocity range 279.6--283.6 km s$^{-1}$, representing the ambient molecular cloud, at a significance level of 3$\sigma$. The position of the point source is marked with a star symbol.}
\label{fig1}

\end{figure}

LMC P3 (also known as 4FGL J0535.2-6736) is a binary system comprising a compact object and an O5 III-type high mass star \citep{2016ApJ...829..105C} and is embedded within the supernova remnant (SNR) DEM L241, located near an H II region in the Large Magellanic Cloud (LMC) \citep{1985ApJS...58..197M,2006A&A...450..585B}. Figure \ref{fig1}(a) shows a large-scale view of the region including DEM L241 and LMC P3, where distributions of Chandra X-rays, H$\alpha$ and S [{\sc ii}] emission using Dark Energy Camera Magellanic Clouds Emission Line Surveys (DeMCELS), and Atacama Submillimeter Telescope Experiment (ASTE) $^{12}$CO($J$=3--2) emission are superposed. The CO cloud is likely the parent molecular cloud which formed the binary LMC P3. The asymmetric X-ray structure extended in the north is peculiar for an SNR and possibly due to past jet activity from the compact object \citep{2012ApJ...759..123S}. Six $\gamma$-ray binaries have been found to date, and LMC P3 is the brightest $\gamma$-ray binary among the six based on Fermi observations in the $<$100 GeV band and H.E.S.S. observations in the 10 TeV band \citep{2016ApJ...829..105C,2017ICRC...35..730K}. It is the first $\gamma$-ray binary discovered outside the Milky Way.

The $\gamma$-ray emission is either due to particle acceleration in the shock from the winds of a compact object and a high mass star companion, or in the relativistic jet driven by mass accretion on the disk around the compact object from the high mass star \citep{2016ApJ...829..105C,2018A&A...610L..17H}. The latter is called a microquasar, a tiny analog to the active galactic nuclei which are much more massive and energetic. Although \cite{2006A&A...450..585B} could not detected coherent pulsation from this source, \cite{2012ApJ...759..123S} found time variability from this source in the X-ray band and this is the first suggestion that this source is a high mass X-ray binary. \cite{2017ICRC...35..730K} searched for periodic emission of LMC P3 in the Fermi-LAT data and found that the MeV/GeV $\gamma$-ray signal is periodic with a period of 10.301 days. Further, the $\gamma$-ray luminosity in an energy range from 200 MeV to 100 GeV is as high as $10^{36}$ erg s$^{-1}$, and the X-ray and radio emission of this object is out of phase with the $\gamma$-ray emission, constraining the $\gamma$ emitting region to be close to the binary system \citep{2016ApJ...829..105C}. LMC P3 is not identified as a microquasar although the binary system is similar to the central engine of a microquasar. We therefore consider that LMC P3 deserves a further pursuit as a candidate of a microquasar, and it is an important task to pursue evidence for an accretion disk around the compact object. 

The origin of cosmic rays (CRs) has been a fundamental issue in astrophysics over several decades, and $\gamma$ rays, a direct signature of CRs, have been an important tool to diagnose CR acceleration. Below the knee at $10^{15.5}$ eV, CRs are thought to be confined within the Galaxy, and SNRs are considered as the most promising candidate of CR accelerators. Recent works established CR acceleration in two TeV $\gamma-$ray SNRs RX J1713.7$-$3946 and RX J0852.0$-$4622, where a significant fraction of $\gamma$ rays were determined to be as the hadronic origin \citep{2021ApJ...915...84F,2024ApJ...961..162F}. It becomes then an issue of common interest if any other population of very high energy objects can be additional CR accelerators. Young massive clusters having high velocity winds are thought to  potentially be a source of CRs (e.g., \citealt{2024NatAs...8..530P}). Most recently, microquasars attracted keen interest as another CR accelerator since they can accelerate charged particles over a wide energy range relevant. At present some 20 microquasars are known in the Milky Way as cataloged by the \cite{2025NSRev..12af496L}, where SS 433, V4641 Sgr, GRS 1915$+$105, MAXI J1820$+$070, and Cygnus X-1 constitute a microquasar sample possibly in the PeV range. These sources are however mostly discovered at optical, radio, and X-rays wavelengths toward directions with small extinctions at latitude $>$2 degrees from the Galactic plane. Most recently, a new methodology to search for microquasars in CO emission was presented for the bright $\gamma-$ray source HESS J1023-575 (HESS J1023 hereafter) which accompanies unique jet and arc CO clouds located in the Galactic midplane at b=0$^{\circ}$ \citep{2009PASJ...61L..23F}. In this object, \cite{fukui2026almaviewjetarcclouds} presented compelling evidence for the footprints of microquasar jets engraved in the CO filamentary clouds with Atacama Large Millimeter/submillimeter Array (ALMA) and suggested its long Myr-term CR acceleration. These authors thus demonstrated CO emission as a tool to diagnose microquasar jets and argued that a microquasar has a potential to be a powerful CR accelerator equivalent to a SNR.

In the present Letter, we investigated the CO emission toward LMC P3 by using the ALMA Atacama Compact Array (ACA) at 0.5 pc resolution and discovered CO jet feature in this region. Our results suggest that such jet-driven molecular structures may be a more common manifestation of jet–ISM interactions than previously recognized. Here we report the first results of this work and discuss their implications, in particular, the high energy activity of the $\gamma-$ray source.

\section{Results}

\begin{figure}
\includegraphics[width=0.9\textwidth]{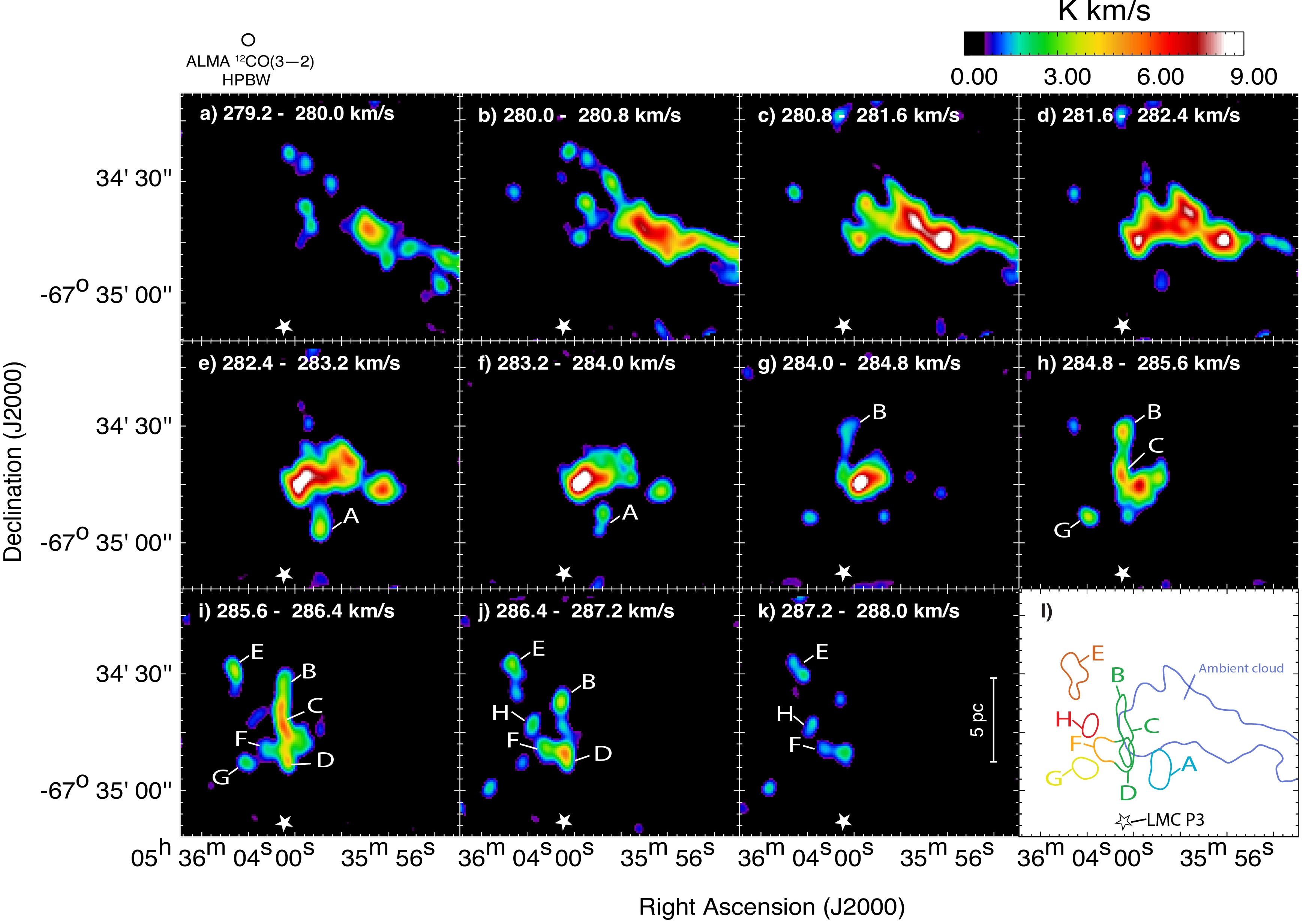}
\caption{Panels (a)$-$(k) present the velocity channel maps from the ALMA $^{12}$CO ($J$=3--2) observations, with a velocity interval of 0.8 km s$^{-1}$. Panel (l) shows the spatial distribution of all identified clouds, displayed as contours in different colors.}
\label{fig2}
\end{figure}

Figure \ref{fig1}(b) shows the distribution of the ALMA ACA $^{12}$CO($J$=3--2) integrated intensity observed in the yellow box in Figure \ref{fig1}(a). The ambient cloud is located at $\sim$10 pc north of LMC P3 which accompanies a thin CO feature with a north-to-south elongation. In order to disentangle the velocity distribution of the CO features, we present velocity channel distributions every 0.8 km s$^{-1}$ for a velocity range from 279.2 km s$^{-1}$ to 288.0 km s$^{-1}$ in Figure \ref{fig2}. The ambient non-jet CO cloud has a velocity range from 280.0 km s$^{-1}$ to 285.6 km s$^{-1}$, and CO jet-like feature is red-shifted in a range from 284.0 km s$^{-1}$ to 288.0 km$^{-1}$. Eight intensity peaks in the red-shifted velocity range are labeled from A through H in each panel of Figure \ref{fig2} in the order of increasing velocity, and constitute a CO jet-like feature and several smaller CO clouds that may share a property similar to the jet-like feature; the CO peaks are grouped into four, i.e., [BCD], [EHF], [A], and [G], according to their locations as summarized in panel (l) of Figure \ref{fig2}. Some small CO clouds in the same velocity range with the ambient cloud are not considered as a CO jet-like feature. The most outstanding feature [BCD] is 8 pc long with width of 1 pc, and shows a remarkable alignment with LMC P3. [EHF] maybe a second jet-like feature consisting of three separated components and is on a line aligned with LMC P3 at an angle of $\sim 16^{\circ}$ to [BCD].  Their length and width are similar to [BCD] in a velocity range of 285.6--288.0 km s$^{-1}$. 

\begin{figure}
\includegraphics[width=0.9\textwidth]{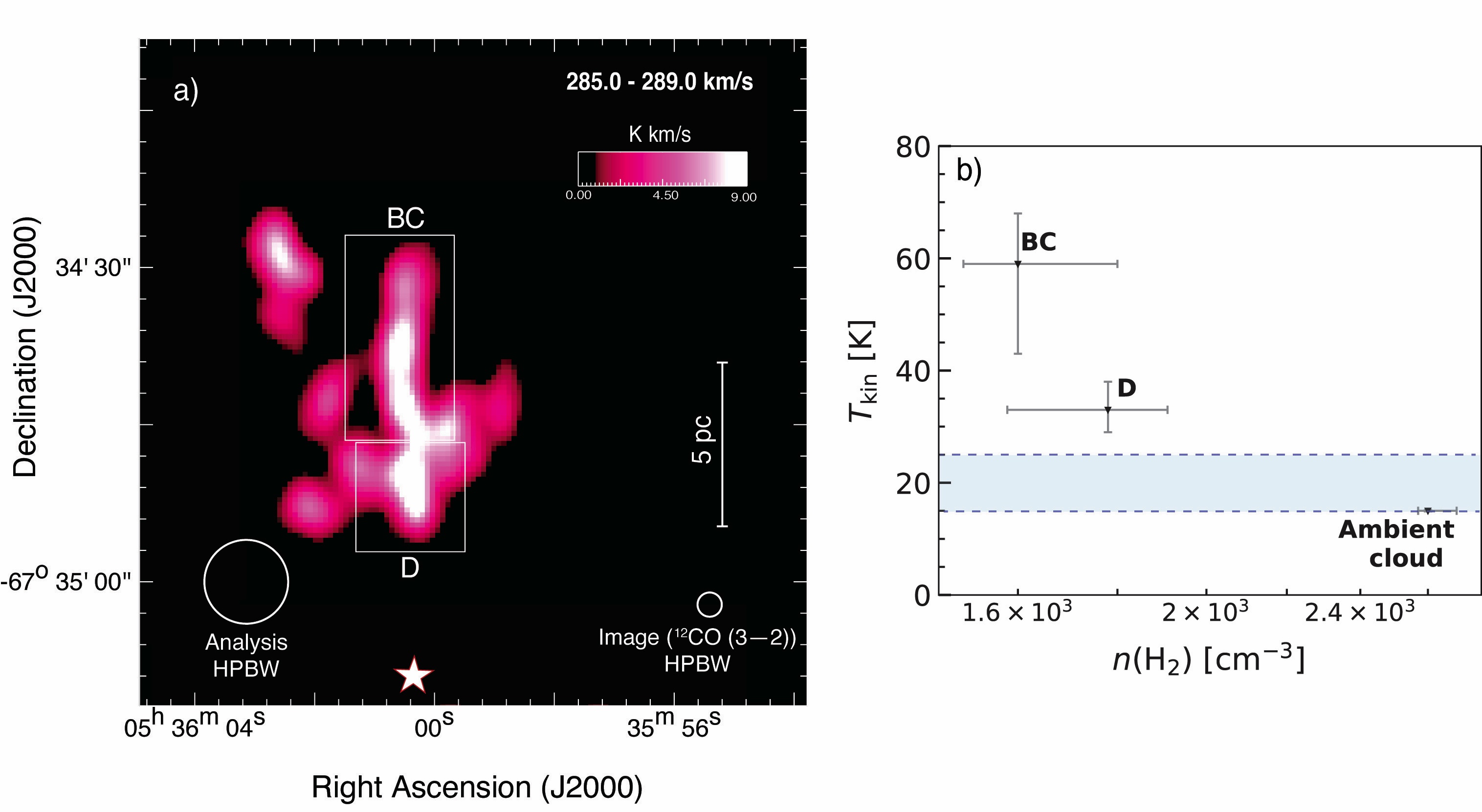}
\caption{a) Intensity map of jet clouds in velocity range 285$-$289 km s$^{-1}$. The boxes indicate the pixel regions selected for estimating physical properties. b) Kinetic temperature of jet clouds BC and D along with ambient cloud estimated in this study are plotted against their density. The horizontal dotted lines indicate the typical temperature range of unshocked cloud in LMC including the uncertainties.}
\label{fig4}
\end{figure}

We derived kinetic temperature of the CO features by applying the Large Velocity Gradient (LVG) approximation toward two bright positions of the jet, BC and D, based on $^{12}$CO($J$=3--2), $^{12}$CO($J$=2--1) and $^{13}$CO($J$=2--1) transitions. The observations used for the analysis is described in detail in Appendix A. We assume that the CO to $H_{2}$ abundance ratio is [$^{12}$CO]/[H$_{2}$]$=1.6\times10^{-5}$ and [$^{12}$CO]/[$^{13}$CO] isotope abundance ratio as 50 \citep{1987ApJ...315..621B,2010PASJ...62...51M,2011AJ....141...73M}. All line data were convolved to a common angular resolution of $\sim8^{\prime\prime}$. At this resolution, sources B and C are not spatially resolved and are therefore treated as a single clump in the analysis. High temperature $\sim$60 K is found in D, and BC shows temperature higher than 30 K. This is significantly higher than $\sim15$ K in the ambient cloud that is obtained by applying the LVG analysis (Figure \ref{fig4}). The typical temperature of the CO clouds without extra-heating in the LMC is  15$-$16 K \citep{2010PASJ...62...51M,2011AJ....141...73M}.

% Please add the following required packages to your document preamble:
% \usepackage{multirow}
% \usepackage[table,xcdraw]{xcolor}
% Beamer presentation requires \usepackage{colortbl} instead of \usepackage[table,xcdraw]{xcolor}
% Please add the following required packages to your document preamble:
% \usepackage{multirow}
%\begin{table}[H]

% Please add the following required packages to your document preamble:
% \usepackage{multirow}

\section{Discussion}

%\subsection{The formation of the Jet-Clouds }

The jet-like cloud [BCD] exhibits an elongated straight shape aligned with LMC P3, and shows significantly increased kinetic temperatures 33$-$60 K as compared with much lower temperatures of the ambient cloud. Considering the location of the $\gamma-$ray source LMC P3, we present a scenario that a high energy jet launched from LMC P3 interacted with the ambient cloud, and formed the jet clouds and heated them up. The present work has revealed that LMC P3 is launching high energy jet which is interacting with the ambient ISM similarly to the microquasar SS433 \citep{2026arXiv260608931S}. Given the binary nature of LMC P3, it is very likely that the mass accretion from the O star to the compact object is feeding the accretion disk and driving the high energy jet. The jet cloud formation is predicted theoretically by the magneto-hydrodynamical simulations of the interaction between microquasar jets with the ambient H {\sc i} gas \citep{2012ApJ...759...35I,2014ApJ...789...79A,2017ApJ...840...25A}. The H {\sc i} gas consists of the cold neutral medium (CNM) and warm neutral medium (WNM), where the CNM clumps are embedded in the extended diffuse WNM.  The microquasar jets compress and heat-up the H {\sc i} gas by a cylindrical shock front which produces a pc-scale compressed layer. In the present case, the ambient gas is a CO cloud enveloped by lower density H {\sc i} gas which is likely responsible for the soft X-ray absorption around the ambient cloud (Figure \ref{fig1}(a)) as discussed below. We therefore assume that the interacting gas is H$_{2}$ gas instead of H {\sc i} gas. As a result, we expect formation of a cylindrical CO gas wall of a pc scale diameter, which we observe as heated-up jet-like CO clouds elongated along the microquasar jet. In this model, we can explain the present jet cloud formation in the CO gas. While the density of the H {\sc i} CNM is 100 cm$^{-3}$ \citep{2017ApJ...840...25A}, about a factor of 10 lower than the CO gas, a similar process will take place also in the CO gas under the extreme high pressure of the microquasar jet. The time scale of the interaction is roughly estimated to be in the order of $3\times10^5$ yr at the jet propagation velocity 0.01 c and for the CO-jet length 8 pc. \cite{2006A&A...450..585B} measured that the X-ray plasma in this SNR is old enough to achieve ionization equilibrium, hence consistent with the present results. The other CO clouds having similar red-shifted velocity as BCD are possibly similar jet-like clouds, which were formed by the putative microquasar jet if it is in precession. This possibility will be explored by follow-up high-resolution ALMA observations in sub-pc scale.

\subsection{High temperature in jet-like clouds}

Concerning the heating in the jet clouds, we find no nearby radiative heating source such as high-mass stars towards the warm regions. The high temperature of jet clouds is thus likely due to the shock heating by the above interaction with the microquasar jet. We see a trend that the northern position BC show temperature higher than the southern cloud D. The higher temperature can be due to more recent heating in the north for a cooling time scale of $10^4$ yr \citep{2000ApJ...532..980K}. The temperatures derived in the ambient cloud are consistent with those of quiescent, unshocked molecular clouds in the LMC \citep{2010PASJ...62...51M,2011AJ....141...73M}. Further discussion on details is due when temperature is estimated at higher angular resolutions.

\subsection{The H {\sc i} envelope of the ambient cloud}

The northern and southern ends of [BCD] are not overlapped with the ambient CO cloud, while the middle part of [BCD] in between shows a similar distribution with the ambient CO cloud with a redshift of 3$-$4 km s$^{-1}$ (Figure \ref{fig1}b). Their distributions suggest that the ambient CO cloud has a more extended lower density gas toward the CO jet which is likely H {\sc i} enveloping the main CO cloud. The two ends of [BCD] were likely formed from H {\sc i} gas as suggested by the X-ray absorption in Figure \ref{fig1}a as argued in Section 3.4. This suggests that the jet-like CO cloud formation occurred also toward the lower density H {\sc i} envelope outside the ambient CO cloud. 

Despite of the acceleration along the jet axis, we find no significant spatial shift of the formed CO jets from the ambient cloud. Such a spatial shift is roughly estimated to be less than 1 pc, if we assume an average velocity by the acceleration of 4 km s$^{-1}$ for the duration less than an age of the SNR (0.8$–$1.6)$\times10^5$ yr \citep{2012ApJ...759..123S}. 

%\subsection{Comparison with Numerical Simulation}

\subsection{A Large Scale View}

The CO jet clouds show a north-biased asymmetry with respect to LMC P3. This asymmetry may have been caused by the CO ambient cloud pre-existent in the north since the CO jet formation requires interstellar gas. If so, the putative microquasar jets can be symmetric both in the north and the south. Although no significant CO emission is detected in the ASTE observation towards the southern region, which is not covered by the current ALMA mosaic (see Figure \ref{fig1}), the higher sensitivity and angular resolution of ALMA may enable the detection of any CO emission, if present. We derive a $3\sigma$ upper limit of $\sim$1.1 K km s$^{-1}$ for the CO emission in this region from the ASTE data. 

\begin{figure}
    \centering
    \includegraphics[width=0.8\linewidth]{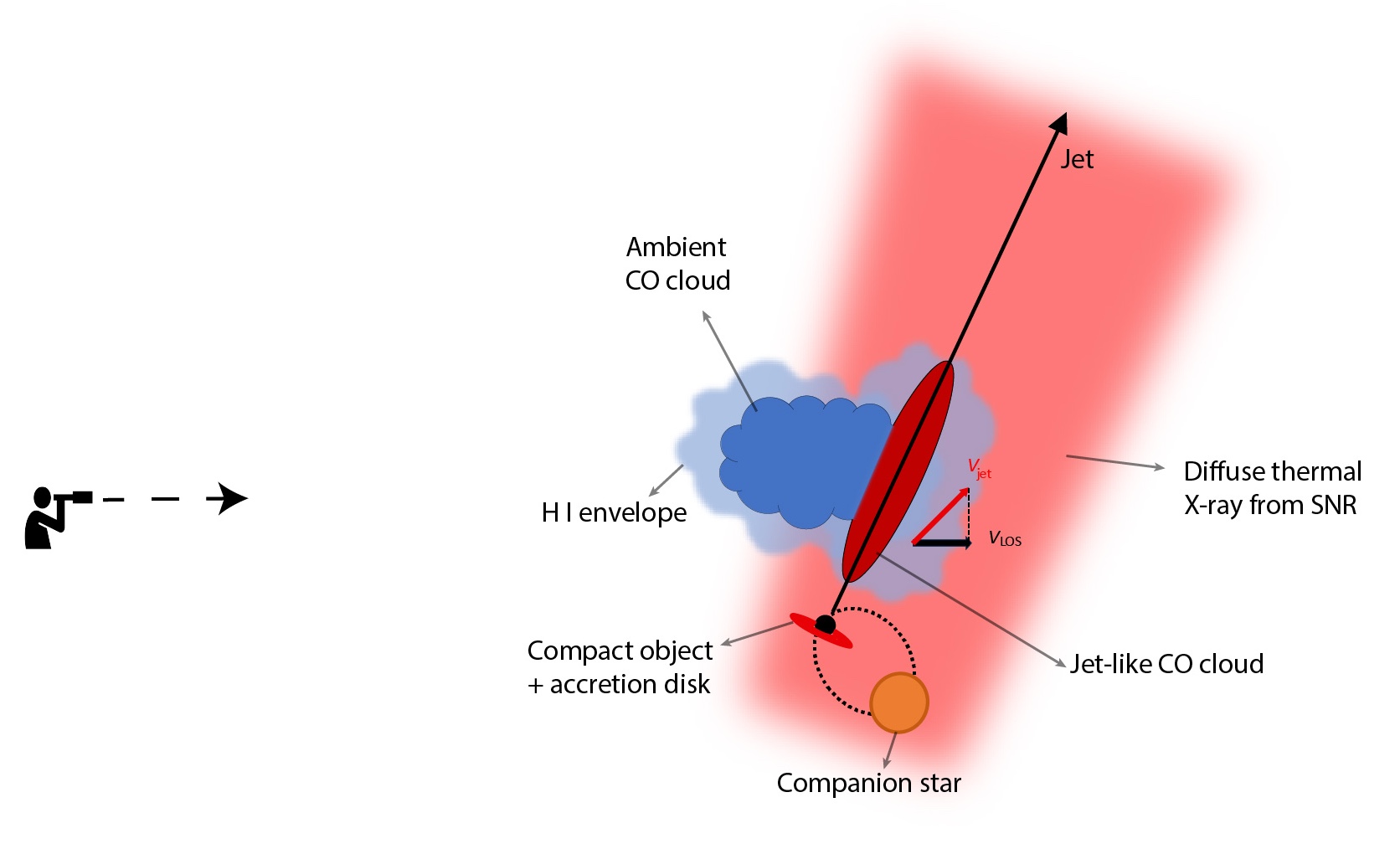}
    \caption{Schematic side-view illustration of the large-scale geometry of the binary system and detected CO clouds within the SNR. The main molecular cloud is located in the foreground of the SNR and is surrounded by an envelope of H I gas. Interaction between the jet launched from the binary system and the ambient gas associated with this envelope is proposed to have produced the jet-like CO cloud located behind the main cloud. }
    \label{fig5}
\end{figure}

The detected CO jet cloud red-shifted relative to the ambient cloud is likely located in the far side of LMC P3 as shown by a side view schematically in Figure \ref{fig5}. We find that a shadow in the X-ray lobe toward 10 pc north coincides with the ambient cloud as is consistent with Figure \ref{fig1}. The shadow is explained by absorption of the soft X-rays when the ambient cloud is located on the near side of the X-ray lobe. We estimate the X-ray absorption column density to be $\sim 1\times10^{22}$ cm$^{-2}$ from the X-ray spectra (Bhuvana et al. 2026, In preparation), which is consistent with the CO gas column density of the ambient cloud. The column density toward the envelope region, corresponding to the shadowed area surrounding the ambient cloud, is found to be $\sim10^{21}$ cm$^{-2}$.

\section{Concluding remarks}
The $\gamma-$ray luminosity of LMC P3 is $\sim10^{36}$ erg s$^{-1}$ \citep{2016ApJ...829..105C} with two potential mechanisms proposed to explain its energy source; one is wind-wind interactions between the compact object and the O star, and the other is mass accretion onto the compact object from the O star winds. In this scenario, the O-type star have a mass loss through their stellar winds of at least $10^{-6}$ M$_{\odot}$ year$^{-1}$, and the released energy corresponds to $6\times10^{40}$ erg s$^{-1}$. Therefore, a fraction of about $10^{-4}$ of the accretion power needs to be converted into $\gamma$-rays, which provides a reasonable supply of the $\gamma-$ray luminosity.
The discovery of the CO jet in LMC P3 has provided strong evidence that LMC P3 is a microquasar launching high energy jets.  This may represent the second case in which continuous CO jet is formed by high energy jets, whereas the high energy jets are not directly detected by high energy radiation, either of the radio synchrotron emission, X-rays, or $\gamma$ rays. Nonetheless, the central object, the binary, is currently releasing large energy as $\gamma$ rays, and is likely accelerating high energy jets and CRs. The non-detection of the high energy jets is consistent with the modest sensitivity of the existing observations, as discussed in \cite{2014ApJ...781...70F}. 
While jet–ISM interactions associated with microquasars have been reported in several systems, formation of the continuous CO jet-like molecular clouds by the interaction was suggested only in HESS J1023$-$575 previously \citep{2009PASJ...61L..23F,2014ApJ...781...70F}. It is obvious that formation of the CO jet requires dense ISM distributed within the Galactic midplane, and LMC P3 is close to the star-forming region N59 having rich ISM in the LMC disk (see \citealt{2019ApJ...885...50W}). HESS J1023$-$575 is also embedded in the mid-plane of the Galaxy with rich ISM as is consistent with the present scenario. On the other hand, the SS433 region is at Galactic latitude of 2 degrees, where the H {\sc i} density is not high \citep{2008PASJ...60..715Y}. The low density is probably a reason why only clumpy CO clouds without a continuous CO jet were formed several 10 pc away from SS433 \citep{2008PASJ...60..715Y,2026arXiv260608931S}. Another microquasar V4641 Sgr located at latitude of 4 degrees, etc., where no CO jet or CO clumps are observed \citep{2026A&A...706A...8A}, probably due to too low ISM density. We therefore argue that extensive searches for CO jet features in the dense ISM environments will be crucial for discovering more microquasar candidates in the Milky Way and the LMC. 

\begin{acknowledgements}
This paper makes use of the following ALMA data: ADS/JAO.ALMA \#2021.2.00008.S. ALMA is a partnership of ESO (representing its member states), NSF (USA) and NINS (Japan), together with NRC (Canada), NSTC and ASIAA (Taiwan), and KASI (Republic of Korea), in cooperation with the Republic of Chile. The Joint ALMA Observatory is operated by ESO, AUI/NRAO and NAOJ. This work was also supported by JSPS KAKENHI grant Nos. 25K17435 (K. Tsuge), 21H01136 (HS) and 24H00246 (HS). This work was supported by the Tokai Pathways to Global Excellence (T-GEx), part of the MEXT Strategic Professional Development Program for Young Researchers, and by the establishment of university fellowships toward the creation of science technology innovation (Grant Number: JPMJFS2138). Support for C.J.L. was provided by NASA through the NASA Hubble Fellowship grant No. HST-HF2-51535.001-A awarded by the Space Telescope Science Institute, which is operated by the Association of Universities for Research in Astronomy, Inc., for NASA, under contract NAS5-26555.
\end{acknowledgements}

\begin{contribution}
%%This section gives authors the space to recognize author contributions. The text inside this environment is NOT counted towards the total word quanta. At a minimum, manuscripts are expected to include this text:

%% But authors are expected to provide more specific details, e.g. 
%%
%%SC was responsible for writing and submitting the manuscript.
%%WWM came up with the initial research concept and edited the manuscript.
%%OTS obtained the funding and edited the manuscript.
%%EBF provided the formal analysis and validation. He also edited the manuscript.
%%GEH Supervised the undergraduates, wrote the software and administers the project github and Zenodo repositories.
%%
%% Authors can use the Contributor Role Taxonomy (CRediT) at
%% https://credit.niso.org
%% for ideas on how write a good statement tailored to their needs.

\end{contribution}

%% To help institutions obtain information on the effectiveness of their 
%% telescopes the AAS Journals has created a group of keywords for telescope 
%% facilities.
%
%% Following the acknowledgments section, use the following syntax and the
%% \facility{} or \facilities{} macros to list the keywords of facilities used 
%% in the research for the paper.  Each keyword is check against the master 
%% list during copy editing.  Individual instruments can be provided in 
%% parentheses, after the keyword, but they are not verified.
\facilities{Chandra, ALMA, ASTE, MOPRA, CTIO}

%% Similar to \facility{}, there is the optional \software command to allow 
%% authors a place to specify which programs were used during the creation of 
%% the manuscript. Authors should list each code and include either a
%% citation or url to the code inside ( )s when available.
\software{CASA \citep{CASA_2022PASP}, IDL Astronomy User’s Library \citep{1995ASPC...77..437L}}

%% Appendix material should be preceded with a single \appendix command.
%% There should be a \section command for each appendix. Mark appendix
%% subsections with the same markup you use in the main body of the paper.
%%
%% Each Appendix (indicated with \section) will be lettered A, B, C, etc.
%% The equation counter will reset when it encounters the \appendix
%% command and will number appendix equations (A1), (A2), etc. The
%% Figure and Table counter will not reset.

\appendix

\section{Observation and Data Reduction} \label{sec:cite}

%\subsection{CO Observations}

We carried out multi-J CO line observations of the SNR DEML 241 using ALMA ACA Band 6 and Band 7 in Cycle 8 (PI: Hidetoshi Sano, \#2021.2.00008.S). The Band 7 observations were taken on 2022, August 22, 27, 30, September 02 and using 11 antennas. The Band 6 data were observed on 2022, August 28, September 08, and 10 using 11 antennas. Further details regarding the observation is provided in \cite{2023ApJ...958...53S}. We also make use of ALMA Total Power (TP) observations carried out on 2022, July 05 for Band 6 and 2022 August 02, 27 for band 7. The observation is centered at $( \alpha_{\mathrm{ICRS}} = 05^{\mathrm{h}}36^{\mathrm{m}}02.79^{\mathrm{s}},\ \delta_{\mathrm{ICRS}} = -67^\circ 34\arcmin 50.57\arcsec)$. Reduction and imaging of observed data were performed using the Common Astronomy Software Applications package (CASA, version 6.5.6-22,\cite{CASA_2022PASP}). We applied its multiscale task \texttt{tclean} with the natural weighting scheme to obtain CLEAN images. The CLEAN 7 m data were then combined with the calibrated TP data using the \texttt{feather} task. The final beam size of the feathered image was $3.3\arcsec\times2.57\arcsec$ with a position angle of $7.72^{\circ}$ for the $^{12}$\text{CO}($J$=3--2) emission line; $7.65{\arcsec}\times5.7{\arcsec}$ with a position angle of $7.12^{\circ}$ for the $^{12}$CO($J$=2--1) emission line; and
$7.62{\arcsec}\times5.66{\arcsec}$ with a position angle of $7.11^{\circ}$ for the $^{13}$\text{CO}($J$=2--1) emission line. The typical rms noise was
$\sim0.07$ K at a velocity resolution of 0.2 km s$^{-1}$ for $^{12}$CO($J$=3--2); $\sim0.07$ K at a velocity resolution of 0.4 km s$^{-1}$ for $^{12}$CO($J$=2--1) and $\sim0.08$ K at a velocity resolution of 0.4 km s$^{-1}$ for $^{13}$CO($J$=2--1) emission line. 

We also make use of $^{12}$CO($J$=1--0) line intensity data from MAGMA survey, obtained with MOPRA telescope covering the region observed by ALMA \citep{2011ApJS..197...16W,2017ApJ...850..139W}. The beam size of MORPA observation is $\sim46\arcsec$. Further, $^{12}$CO($J$=3--2) line intensity observation of the whole SNR DEM L241 made by ASTE is used. The data were obtained as part of the ASTE CO(3--2) survey of LMC supernova remnants under Proposal ID AC141023 (see \cite{2018ApJ...867....7S,2019ApJ...873...40S} for furter details).

We utilized archival Chandra X-ray observation of DEM L241 contained in the Chandra Data Collection\dataset[DOI: https://doi.org/10.25574/cdc.649]{https://doi.org/10.25574/cdc.649} (observation IDL: 13226, PI: Seward). The observation was carried out on 2011 February 8 using the Advanced CCD Imaging Spectrometer (ACIS). Since only imaging was required, the standard pipeline–processed data provided by the Chandra X-ray Center (CXC) were used without reprocessing. The image was created in 0.3$-$10 keV from the Level-2 event file using the CIAO software package (version  4.17.0) and CALDB (version 4.11.6).

Optical images of DEM L241 were obtained from the Dark Energy Camera Magellanic Clouds Emission Line Survey (DeMCELS) \citep{2024ApJ...974...70P}. We used publicly available (DR1) narrowband images in H$\alpha$ and [S {\sc ii}] bands. These images were acquired using the Dark Energy Camera (DECam) mounted on the Blanco 4-m telescope at CTIO, as part of the DeMCELS program \citep{2015AJ....150..150F}.

\bibliography{sample701}

@ARTICLE{2021ApJ...915...84F,
       author = {{Fukui}, Yasuo and {Sano}, Hidetoshi and {Yamane}, Yumiko and {Hayakawa}, Takahiro and {Inoue}, Tsuyoshi and {Tachihara}, Kengo and {Rowell}, Gavin and {Einecke}, Sabrina},
        title = "{Pursuing the Origin of the Gamma Rays in RX J1713.7-3946 Quantifying the Hadronic and Leptonic Components}",
      journal = {\apj},
         year = 2021,
        month = jul,
       volume = {915},
       number = {2},
          eid = {84},
        pages = {84},
          doi = {10.3847/1538-4357/abff4a},
archivePrefix = {arXiv},
       eprint = {2105.02734},
 primaryClass = {astro-ph.HE},
       adsurl = {https://ui.adsabs.harvard.edu/abs/2021ApJ...915...84F}
}

@ARTICLE{2024NatAs...8..530P,
       author = {{Peron}, Giada and {Casanova}, Sabrina and {Gabici}, Stefano and {Baghmanyan}, Vardan and {Aharonian}, Felix},
        title = "{The contribution of winds from star clusters to the Galactic cosmic-ray population}",
      journal = {Nature Astronomy},
         year = 2024,
        month = apr,
       volume = {8},
        pages = {530-537},
          doi = {10.1038/s41550-023-02168-6},
archivePrefix = {arXiv},
       eprint = {2407.07509},
 primaryClass = {astro-ph.HE},
       adsurl = {https://ui.adsabs.harvard.edu/abs/2024NatAs...8..530P}
}

@ARTICLE{2023ApJ...958...53S,
       author = {{Sano}, H. and {Yamane}, Y. and {van Loon}, J. Th. and {Furuya}, K. and {Fukui}, Y. and {Alsaberi}, R.~Z.~E. and {Bamba}, A. and {Enokiya}, R. and {Filipovi{\'c}}, M.~D. and {Indebetouw}, R. and {Inoue}, T. and {Kawamura}, A. and {Laki{\'c}evi{\'c}}, M. and {Law}, C.~J. and {Mizuno}, N. and {Murase}, T. and {Onishi}, T. and {Park}, S. and {Plucinsky}, P.~P. and {Rho}, J. and {Richards}, A.~M.~S. and {Rowell}, G. and {Sasaki}, M. and {Seok}, J. and {Sharda}, P. and {Staveley-Smith}, L. and {Suzuki}, H. and {Temim}, T. and {Tokuda}, K. and {Tsuge}, K. and {Tachihara}, K.},
        title = "{ALMA Observations of Supernova Remnant N49 in the Large Magellanic Cloud. II. Non-LTE Analysis of Shock-heated Molecular Clouds}",
      journal = {\apj},
         year = 2023,
        month = nov,
       volume = {958},
       number = {1},
          eid = {53},
        pages = {53},
          doi = {10.3847/1538-4357/acffbe},
archivePrefix = {arXiv},
       eprint = {2311.02180},
 primaryClass = {astro-ph.GA},
       adsurl = {https://ui.adsabs.harvard.edu/abs/2023ApJ...958...53S}
}

@misc{fukui2026almaviewjetarcclouds,
      title={An ALMA view of the Jet-Arc CO clouds toward the TeV $\gamma$-ray source HESS J1023-575 and Westerlund 2; Evidence for the footprints of microquasar jets, the very powerful cosmic-ray accelerator in the Galactic disk}, 
      author={Yasuo Fukui and Kisetsu Tsuge and Hidetoshi Sano and G. R. Bhuvana and Rin I. Yamada and Ryoji Matsumoto and Yuta Asahina and Tsuyoshi Inoue and Tim Lukas Holch and Emma de Oña Wilhelmi},
      year={2026},
      eprint={2608.14988},
      archivePrefix={arXiv},
      primaryClass={astro-ph.HE},
      url={https://arxiv.org/abs/2608.14988} 
}

@ARTICLE{CASA_2022PASP,
       author = {{Team}, CASA  and {Bean}, Ben and {Bhatnagar}, Sanjay and {Castro}, Sandra and {Donovan Meyer}, Jennifer and {Emonts}, Bjorn and {Garcia}, Enrique and {Garwood}, Robert and {Golap}, Kumar and {Gonzalez Villalba}, Justo and {Harris}, Pamela and {Hayashi}, Yohei and {Hoskins}, Josh and {Hsieh}, Mingyu and {Jagannathan}, Preshanth and {Kawasaki}, Wataru and {Keimpema}, Aard and {Kettenis}, Mark and {Lopez}, Jorge and {Marvil}, Joshua and {Masters}, Joseph and {McNichols}, Andrew and {Mehringer}, David and {Miel}, Renaud and {Moellenbrock}, George and {Montesino}, Federico and {Nakazato}, Takeshi and {Ott}, Juergen and {Petry}, Dirk and {Pokorny}, Martin and {Raba}, Ryan and {Rau}, Urvashi and {Schiebel}, Darrell and {Schweighart}, Neal and {Sekhar}, Srikrishna and {Shimada}, Kazuhiko and {Small}, Des and {Steeb}, Jan-Willem and {Sugimoto}, Kanako and {Suoranta}, Ville and {Tsutsumi}, Takahiro and {van Bemmel}, Ilse M. and {Verkouter}, Marjolein and {Wells}, Akeem and {Xiong}, Wei and {Szomoru}, Arpad and {Griffith}, Morgan and {Glendenning}, Brian and {Kern}, Jeff},
        title = {CASA, the Common Astronomy Software Applications for Radio Astronomy},
      journal = {\pasp},
         year = 2022,
        month = nov,
       volume = {134},
       number = {1041},
          eid = {114501},
        pages = {114501},
          doi = {10.1088/1538-3873/ac9642},
archivePrefix = {arXiv},
       eprint = {2210.02276},
 primaryClass = {astro-ph.IM},
       adsurl = {https://ui.adsabs.harvard.edu/abs/2022PASP..134k4501C}
}

@INPROCEEDINGS{1995ASPC...77..437L,
       author = {{Landsman}, W.~B.},
        title = "{The IDL Astronomy User's Library}",
    booktitle = {Astronomical Data Analysis Software and Systems IV},
         year = 1995,
       editor = {{Shaw}, R.~A. and {Payne}, H.~E. and {Hayes}, J.~J.~E.},
       series = {Astronomical Society of the Pacific Conference Series},
       volume = {77},
        month = jan,
        pages = {437},
       adsurl = {https://ui.adsabs.harvard.edu/abs/1995ASPC...77..437L}
}

@ARTICLE{2025NSRev..12af496L,
       author = {{LHAASO Collaboration} and {Cao}, Zhen and {Aharonian}, Felix and {Bai}, Yun-Xiang and {Bao}, Yi-Wei and {Bastieri}, Denis and {Bi}, Xiao-Jun and {Bi}, Yu-Jiang and {Bian}, Wen-Yi and {Bukevich}, Anatoly V. and {Cai}, Chengmiao and {Cao}, Wen-Yu and {Cao}, Zhe and {Chang}, Jin and {Chang}, Jin-Fan and {Chen}, Aming and {Chen}, En-Sheng and {Chen}, Guohai and {Chen}, Hua-Xi and {Chen}, Liang and {Chen}, Long and {Chen}, Ming-Jun and {Chen}, Ma-Li and {Chen}, Qi-Hui and {Chen}, Shi and {Chen}, Su-Hong and {Chen}, Song-Zhan and {Chen}, Tian-Lu and {Chen}, Xiao-Bin and {Chen}, Xuejian and {Chen}, Yang and {Cheng}, Ning and {Cheng}, Yao-Dong and {Chung Chu}, Ming and {Cui}, Ming-Yang and {Cui}, Shu-Wang and {Cui}, Xiao-Hong and {Cui}, Yi-Dong and {Dai}, Ben-Zhong and {Dai}, Hong-Liang and {Dai}, Zigao and {Luobu}, Danzeng and {Diao}, Yang-Xuan and {Dong}, Xu-Qiang and {Duan}, Kai-Kai and {Fan}, Jun-Hui and {Fan}, Yi-Zhong and {Fang}, Jun and {Fang}, Jian-Hua and {Fang}, Kun and {Feng}, Cun-Feng and {Feng}, Hua and {Feng}, Li and {Feng}, Shaohui and {Feng}, Xiao-Ting and {Feng}, Yi and {Feng}, You-Liang and {Gabici}, Stefano and {Gao}, Bo and {Gao}, Chuan-Dong and {Gao}, Qi and {Gao}, Wei and {Gao}, Wei-Kang and {Ge}, Maomao and {Ge}, Ting-Ting and {Geng}, Lisi and {Giacinti}, Gwenael and {Gong}, Guanghua and {Gou}, Quanbu and {Gu}, Min-Hao and {Guo}, Fu-Lai and {Guo}, Jing and {Guo}, Xiao-Lei and {Guo}, Yi-Qing and {Guo}, Ying-Ying and {Han}, Yi-Ang and {Hannuksela}, Otto A. and {Hasan}, Mariam and {He}, Hui-Hai and {He}, Hao-Ning and {He}, Jia-Yin and {He}, Xinyu and {He}, Yu and {Hern{\'a}ndez-Cadena}, Sergio and {Hou}, Bo-Wen and {Hou}, Chao and {Hou}, Xian and {Hu}, Hong-Bo and {Hu}, Shi-Cong and {Huang}, Chen and {Huang}, Dai-Hui and {Huang}, Jiajun and {Huang}, Tian-Qi and {Huang}, Wen-Jun and {Huang}, Xing-Tao and {Huang}, Xiao-Yuan and {Huang}, Yong and {Huang}, Yi-Yun and {Ji}, Xiao-Lu and {Jia}, Huan-Yu and {Jia}, Kang and {Jiang}, Hou-Bing and {Jiang}, Kun and {Jiang}, Xiao-Wei and {Jiang}, Ze-Jun and {Jin}, Min and {Kaci}, Samy and {Kang}, Ming-Ming and {Karpikov}, Ivan and {Khangulyan}, Dmitry and {Kuleshov}, Denis and {Kurinov}, Kirill and {Li}, Bing-Bing and {Li}, Cheng and {Li}, Cong and {Li}, Dan and {Li}, Fei and {Li}, Haibo and {Li}, Huicai and {Li}, Jian and {Li}, Jie and {Li}, Kai and {Li}, Long and {Li}, Rong-Lan and {Li}, Si-Da and {Li}, Tian-Yang and {Li}, Wen-Lian and {Li}, Xiu-Rong and {Li}, Xin and {Li}, Yuan and {Li}, Yizhuo and {Li}, Zhe and {Li}, Zhuo and {Liang}, En-Wei and {Liang}, Yun-Feng and {Lin}, Su-Jie and {Liu}, Bing and {Liu}, Cheng and {Liu}, Dong and {Liu}, Dang-Bo and {Liu}, Hu and {Liu}, Hai-Dong and {Liu}, Jia and {Liu}, Jia-Li and {Liu}, Ji-Ren and {Liu}, Mao-Yuan and {Liu}, Ruo-Yu and {Liu}, Si-Ming and {Liu}, Wei and {Liu}, X. and {Liu}, Yi and {Liu}, Yu and {Liu}, Yi-Nong and {Lou}, Yu-Qing and {Luo}, Qing and {Luo}, Yu and {Lv}, Hong-Kui and {Ma}, Bo-Qiang and {Ma}, Ling-Ling and {Ma}, Xin-Hua and {Mao}, Ji-Rong and {Min}, Zhen and {Mitthumsiri}, Warit and {Mou}, Guo-Bin and {Mu}, Hui-Jun and {Neronov}, Andrii and {Ng}, Kenny Chun Yu and {Ni}, Ming-Yang and {Nie}, Lin and {Ou}, Le-Jian and {Pattarakijwanich}, Petchara and {Pei}, Zhi-Yuan and {Qi}, Jin-Can and {Qi}, Meng-Yao and {Qin}, Jia-Jun and {Raza}, Ali and {Ren}, Chong-Yang and {Ruffolo}, David and {S{\'a}iz}, Alejandro and {Semikoz}, Dmitri and {Shao}, Lang and {Shchegolev}, Oleg and {Shen}, Yun-Zhi and {Sheng}, Xiang-Dong and {Shi}, Zhaodong and {Shu}, Fu-Wen and {Song}, Hui-Chao and {Stenkin}, Yuri V. and {Stepanov}, Vladimir and {Su}, Yang and {Sun}, Dongxu and {Sun}, Hao and {Sun}, Qinning and {Sun}, Xiaona and {Sun}, Zhibin and {Hussain Tabasam}, Nabeel and {Takata}, Jumpei and {Tam}, Pak Hin Thomas and {Tan}, Hong-Bin and {Tang}, Qingwen},
        title = "{Ultrahigh-Energy Gamma-ray Emission Associated with Black Hole-Jet Systems}",
      journal = {National Science Review},
         year = 2025,
        month = dec,
       volume = {12},
       number = {12},
          eid = {nwaf496},
        pages = {nwaf496},
          doi = {10.1093/nsr/nwaf496},
archivePrefix = {arXiv},
       eprint = {2410.08988},
 primaryClass = {astro-ph.HE},
       adsurl = {https://ui.adsabs.harvard.edu/abs/2025NSRev..12af496L}
}

@ARTICLE{2024ApJ...961..162F,
       author = {{Fukui}, Yasuo and {Aruga}, Maki and {Sano}, Hidetoshi and {Hayakawa}, Takahiro and {Inoue}, Tsuyoshi and {Rowell}, Gavin and {Einecke}, Sabrina and {Tachihara}, Kengo},
        title = "{The Gamma-Ray Origin of RX J0852.0-4622 Quantifying the Hadronic and Leptonic Components: Further Evidence for the Cosmic-Ray Acceleration in Young Shell-type SNRs}",
      journal = {\apj},
         year = 2024,
        month = feb,
       volume = {961},
       number = {2},
          eid = {162},
        pages = {162},
          doi = {10.3847/1538-4357/ad0da3},
archivePrefix = {arXiv},
       eprint = {2311.11355},
 primaryClass = {astro-ph.HE},
       adsurl = {https://ui.adsabs.harvard.edu/abs/2024ApJ...961..162F}
}

@ARTICLE{2014ApJ...789...79A,
       author = {{Asahina}, Yuta and {Ogawa}, Takayuki and {Kawashima}, Tomohisa and {Furukawa}, Naoko and {Enokiya}, Rei and {Yamamoto}, Hiroaki and {Fukui}, Yasuo and {Matsumoto}, Ryoji},
        title = "{Magnetohydrodynamic Simulations of a Jet Drilling an H I Cloud: Shock Induced Formation of Molecular Clouds and Jet Breakup}",
      journal = {\apj},
         year = 2014,
        month = jul,
       volume = {789},
       number = {1},
          eid = {79},
        pages = {79},
          doi = {10.1088/0004-637X/789/1/79},
archivePrefix = {arXiv},
       eprint = {1407.2381},
 primaryClass = {astro-ph.HE},
       adsurl = {https://ui.adsabs.harvard.edu/abs/2014ApJ...789...79A}
}

@ARTICLE{2017ApJ...840...25A,
       author = {{Asahina}, Yuta and {Nomura}, Mariko and {Ohsuga}, Ken},
        title = "{Enhancement of Feedback Efficiency by Active Galactic Nucleus Outflows via the Magnetic Tension Force in the Inhomogeneous Interstellar Medium}",
      journal = {\apj},
         year = 2017,
        month = may,
       volume = {840},
       number = {1},
          eid = {25},
        pages = {25},
          doi = {10.3847/1538-4357/aa6c5f},
       adsurl = {https://ui.adsabs.harvard.edu/abs/2017ApJ...840...25A}
}

@ARTICLE{2026A&A...706A...8A,
       author = {{Acharyya}, A. and {Aharonian}, F. and {Ashkar}, H. and {Backes}, M. and {Batzofin}, R. and {Berge}, D. and {Bernl{\"o}hr}, K. and {B{\"o}ttcher}, M. and {Boisson}, C. and {Bolmont}, J. and {Brun}, F. and {Bruno}, B. and {Burger-Scheidlin}, C. and {Bylund}, T. and {Casanova}, S. and {Celic}, J. and {Cerruti}, M. and {Chen}, A. and {Chernyakova}, M. and {Chibueze}, J.~O. and {Chibueze}, O. and {Cornejo}, B. and {Cotter}, G. and {de Assis Scarpin}, J. and {de Bony de Lavergne}, M. and {de Naurois}, M. and {de O{\~n}a Wilhelmi}, E. and {Delgado Giler}, A.~G. and {Devin}, J. and {Djannati-Ata{\"\i}}, A. and {Dmytriiev}, A. and {Egberts}, K. and {Egg}, K. and {Ernenwein}, J.-P. and {Esca{\~n}uela Nieves}, C. and {Fauverge}, P. and {Feijen}, K. and {Filipovic}, M.~D. and {Fontaine}, G. and {Funk}, S. and {Gabici}, S. and {Gallant}, Y.~A. and {Glicenstein}, J.~F. and {Glombitza}, J. and {Goswami}, P. and {Grondin}, M.-H. and {Heckmann}, L. and {He{\ss}}, B. and {Hinton}, J.~A. and {Hofmann}, W. and {Holch}, T.~L. and {Holler}, M. and {Jamrozy}, M. and {Jankowsky}, F. and {Jardin-Blicq}, A. and {Jaroschewski}, I. and {Jimeno}, D. and {Jung-Richardt}, I. and {Katarzy{\'n}ski}, K. and {Kerszberg}, D. and {Kh{\'e}lifi}, B. and {Komin}, N. and {Kosack}, K. and {Kostunin}, D. and {Lang}, R.~G. and {Lazarevi{\'c}}, S. and {Lemi{\`e}re}, A. and {Lemoine-Goumard}, M. and {Lenain}, J.-P. and {Liniewicz}, P. and {Luashvili}, A. and {Mackey}, J. and {Malyshev}, D. and {Marandon}, V. and {Mayer}, M.~G.~F. and {Mehta}, A. and {Mitchell}, A.~M.~W. and {Moderski}, R. and {Mohrmann}, L. and {Montanari}, A. and {Moulin}, E. and {Niemiec}, J. and {Olivera-Nieto}, L. and {Moghadam}, M.~O. and {Panny}, S. and {Parsons}, R.~D. and {Pensec}, U. and {Pichard}, P. and {Preis}, T. and {P{\"u}hlhofer}, G. and {Punch}, M. and {Quirrenbach}, A. and {Reimer}, A. and {Reimer}, O. and {Reis}, I. and {Remy}, Q. and {Ren}, H.~X. and {Reville}, B. and {Rieger}, F. and {Roellinghoff}, G. and {Rowell}, G. and {Rudak}, B. and {Sabri}, K. and {Safi-Harb}, S. and {Sahakian}, V. and {Santangelo}, A. and {Sasaki}, M. and {Sch{\"u}ssler}, F. and {Shapopi}, J.~N.~S. and {Si Said}, W. and {Sol}, H. and {Stawarz}, {\L}. and {Steinmassl}, S. and {Tanaka}, T. and {Taylor}, A.~M. and {Taylor}, G.~L. and {Terrier}, R. and {Tian}, Y. and {Timmermans}, A. and {Tsirou}, M. and {Tsuji}, N. and {Unbehaun}, T. and {van Eldik}, C. and {Vecchi}, M. and {Venter}, C. and {Vink}, J. and {Voitsekhovskyi}, V. and {Wagner}, S.~J. and {Wierzcholska}, A. and {Zacharias}, M. and {Zdziarski}, A.~A. and {Zech}, A. and {Zhong}, W. and {H.~E.~S.~S. Collaboration} and {Takekawa}, S.},
        title = "{Constraining the nature of the most extreme Galactic particle accelerator: H.E.S.S. observations of the microquasar V4641 Sgr}",
      journal = {\aap},
         year = 2026,
        month = jan,
       volume = {706},
          eid = {A8},
        pages = {A8},
          doi = {10.1051/0004-6361/202557532},
archivePrefix = {arXiv},
       eprint = {2511.10537},
 primaryClass = {astro-ph.HE},
       adsurl = {https://ui.adsabs.harvard.edu/abs/2026A&A...706A...8A}
}

@ARTICLE{2011AJ....141...73M,
       author = {{Minamidani}, Tetsuhiro and {Tanaka}, Takanori and {Mizuno}, Yoji and {Mizuno}, Norikazu and {Kawamura}, Akiko and {Onishi}, Toshikazu and {Hasegawa}, Tetsuo and {Tatematsu}, Ken'ichi and {Takekoshi}, Tatsuya and {Sorai}, Kazuo and {Moribe}, Nayuta and {Torii}, Kazufumi and {Sakai}, Takeshi and {Muraoka}, Kazuyuki and {Tanaka}, Kunihiko and {Ezawa}, Hajime and {Kohno}, Kotaro and {Kim}, Sungeun and {Rubio}, M{\'o}nica and {Fukui}, Yasuo},
        title = "{Dense Clumps in Giant Molecular Clouds in the Large Magellanic Cloud: Density and Temperature Derived from $^{13}$CO(J = 3-2) Observations}",
      journal = {\aj},
         year = 2011,
        month = mar,
       volume = {141},
       number = {3},
          eid = {73},
        pages = {73},
          doi = {10.1088/0004-6256/141/3/73},
archivePrefix = {arXiv},
       eprint = {1012.5037},
 primaryClass = {astro-ph.GA},
       adsurl = {https://ui.adsabs.harvard.edu/abs/2011AJ....141...73M}
}

@ARTICLE{2024ApJ...974...70P,
       author = {{Points}, Sean D. and {Long}, Knox S. and {Blair}, William P. and {Williams}, Rosa and {Chu}, You-Hua and {Winkler}, P. Frank and {White}, Richard L. and {Rest}, Armin and {Li}, Chuan-Jui and {Valdes}, Francisco},
        title = "{The Dark Energy Camera Magellanic Clouds Emission-line Survey}",
      journal = {\apj},
         year = 2024,
        month = oct,
       volume = {974},
       number = {1},
          eid = {70},
        pages = {70},
          doi = {10.3847/1538-4357/ad6766},
archivePrefix = {arXiv},
       eprint = {2409.04846},
 primaryClass = {astro-ph.GA},
       adsurl = {https://ui.adsabs.harvard.edu/abs/2024ApJ...974...70P}
}

@ARTICLE{2015AJ....150..150F,
       author = {{Flaugher}, B. and {Diehl}, H.~T. and {Honscheid}, K. and {Abbott}, T.~M.~C. and {Alvarez}, O. and {Angstadt}, R. and {Annis}, J.~T. and {Antonik}, M. and {Ballester}, O. and {Beaufore}, L. and {Bernstein}, G.~M. and {Bernstein}, R.~A. and {Bigelow}, B. and {Bonati}, M. and {Boprie}, D. and {Brooks}, D. and {Buckley-Geer}, E.~J. and {Campa}, J. and {Cardiel-Sas}, L. and {Castander}, F.~J. and {Castilla}, J. and {Cease}, H. and {Cela-Ruiz}, J.~M. and {Chappa}, S. and {Chi}, E. and {Cooper}, C. and {da Costa}, L.~N. and {Dede}, E. and {Derylo}, G. and {DePoy}, D.~L. and {de Vicente}, J. and {Doel}, P. and {Drlica-Wagner}, A. and {Eiting}, J. and {Elliott}, A.~E. and {Emes}, J. and {Estrada}, J. and {Fausti Neto}, A. and {Finley}, D.~A. and {Flores}, R. and {Frieman}, J. and {Gerdes}, D. and {Gladders}, M.~D. and {Gregory}, B. and {Gutierrez}, G.~R. and {Hao}, J. and {Holland}, S.~E. and {Holm}, S. and {Huffman}, D. and {Jackson}, C. and {James}, D.~J. and {Jonas}, M. and {Karcher}, A. and {Karliner}, I. and {Kent}, S. and {Kessler}, R. and {Kozlovsky}, M. and {Kron}, R.~G. and {Kubik}, D. and {Kuehn}, K. and {Kuhlmann}, S. and {Kuk}, K. and {Lahav}, O. and {Lathrop}, A. and {Lee}, J. and {Levi}, M.~E. and {Lewis}, P. and {Li}, T.~S. and {Mandrichenko}, I. and {Marshall}, J.~L. and {Martinez}, G. and {Merritt}, K.~W. and {Miquel}, R. and {Mu{\~n}oz}, F. and {Neilsen}, E.~H. and {Nichol}, R.~C. and {Nord}, B. and {Ogando}, R. and {Olsen}, J. and {Palaio}, N. and {Patton}, K. and {Peoples}, J. and {Plazas}, A.~A. and {Rauch}, J. and {Reil}, K. and {Rheault}, J.-P. and {Roe}, N.~A. and {Rogers}, H. and {Roodman}, A. and {Sanchez}, E. and {Scarpine}, V. and {Schindler}, R.~H. and {Schmidt}, R. and {Schmitt}, R. and {Schubnell}, M. and {Schultz}, K. and {Schurter}, P. and {Scott}, L. and {Serrano}, S. and {Shaw}, T.~M. and {Smith}, R.~C. and {Soares-Santos}, M. and {Stefanik}, A. and {Stuermer}, W. and {Suchyta}, E. and {Sypniewski}, A. and {Tarle}, G. and {Thaler}, J. and {Tighe}, R. and {Tran}, C. and {Tucker}, D. and {Walker}, A.~R. and {Wang}, G. and {Watson}, M. and {Weaverdyck}, C. and {Wester}, W. and {Woods}, R. and {Yanny}, B. and {DES Collaboration}},
        title = "{The Dark Energy Camera}",
      journal = {\aj},
         year = 2015,
        month = nov,
       volume = {150},
       number = {5},
          eid = {150},
        pages = {150},
          doi = {10.1088/0004-6256/150/5/150},
archivePrefix = {arXiv},
       eprint = {1504.02900},
 primaryClass = {astro-ph.IM},
       adsurl = {https://ui.adsabs.harvard.edu/abs/2015AJ....150..150F}
}

@ARTICLE{1987ApJ...315..621B,
       author = {{Blake}, Geoffrey A. and {Sutton}, E.~C. and {Masson}, C.~R. and {Phillips}, T.~G.},
        title = "{Molecular Abundances in OMC-1: The Chemical Composition of Interstellar Molecular Clouds and the Influence of Massive Star Formation}",
      journal = {\apj},
         year = 1987,
        month = apr,
       volume = {315},
        pages = {621},
          doi = {10.1086/165165},
       adsurl = {https://ui.adsabs.harvard.edu/abs/1987ApJ...315..621B}
}

@ARTICLE{2009PASJ...61L..23F,
       author = {{Fukui}, Yasuo and {Furukawa}, Naoko and {Dame}, Thomas M. and {Dawson}, Joanne R. and {Yamamoto}, Hiroaki and {Rowell}, Gavin P. and {Aharonian}, Felix and {Hofmann}, Werner and {de O{\~n}a Wilhelmi}, Emma and {Minamidani}, Tetsuhiro and {Kawamura}, Akiko and {Mizuno}, Norikazu and {Onishi}, Toshikazu and {Mizuno}, Akira and {Nagataki}, Shigehiro},
        title = "{A Peculiar Jet and Arc of Molecular Gas toward the Rich and Young Stellar Cluster Westerlund 2 and a TeV Gamma Ray Source}",
      journal = {\pasj},
         year = 2009,
        month = aug,
       volume = {61},
        pages = {L23},
          doi = {10.1093/pasj/61.4.L23},
archivePrefix = {arXiv},
       eprint = {0903.5340},
 primaryClass = {astro-ph.HE},
       adsurl = {https://ui.adsabs.harvard.edu/abs/2009PASJ...61L..23F}
}

@ARTICLE{2008PASJ...60..715Y,
       author = {{Yamamoto}, Hiroaki and {Ito}, Shingo and {Ishigami}, Shinji and {Fujishita}, Motosuji and {Kawase}, Tokuichi and {Kawamura}, Akiko and {Mizuno}, Norikazu and {Onishi}, Toshikazu and {Mizuno}, Akira and {McClure-Griffiths}, Naomi M. and {Fukui}, Yasuo},
        title = "{Aligned Molecular Clouds towards SS 433 and L = 348{\textdegree}.5: Possible Evidence for a Galactic ``Vapor Trail'' Created by a Relativistic Jet}",
      journal = {\pasj},
         year = 2008,
        month = aug,
       volume = {60},
        pages = {715},
          doi = {10.1093/pasj/60.4.715},
archivePrefix = {arXiv},
       eprint = {0804.1871},
 primaryClass = {astro-ph},
       adsurl = {https://ui.adsabs.harvard.edu/abs/2008PASJ...60..715Y}
}

@ARTICLE{2019ApJ...885...50W,
       author = {{Wong}, Tony and {Hughes}, Annie and {Tokuda}, Kazuki and {Indebetouw}, R{\'e}my and {Onishi}, Toshikazu and {Bandurski}, Jeffrey B. and {Chen}, C.-H. Rosie and {Fukui}, Yasuo and {Glover}, Simon C.~O. and {Klessen}, Ralf S. and {Pineda}, Jorge L. and {Roman-Duval}, Julia and {Sewi{\l}o}, Marta and {Wojciechowski}, Evan and {Zahorecz}, Sarolta},
        title = "{Relations between Molecular Cloud Structure Sizes and Line Widths in the Large Magellanic Cloud}",
      journal = {\apj},
         year = 2019,
        month = nov,
       volume = {885},
       number = {1},
          eid = {50},
        pages = {50},
          doi = {10.3847/1538-4357/ab46ba},
archivePrefix = {arXiv},
       eprint = {1905.11827},
 primaryClass = {astro-ph.GA},
       adsurl = {https://ui.adsabs.harvard.edu/abs/2019ApJ...885...50W}
}

@ARTICLE{2026arXiv260608931S,
       author = {{Sakemi}, Haruka and {Sano}, Hidetoshi and {Fukui}, Yasuo and {Machida}, Mami and {Kimura}, Shigeo S. and {Kobayashi}, Masato I.~N. and {Kayama}, Kazuho and {Yamamoto}, Hiroaki and {Tachihara}, Kengo and {Nagai}, Hiroshi},
        title = "{Discovery of CO Clouds Associated with the X-ray Jets of SS 433: Evidence for Shock-Cloud Interaction Enhancing Nonthermal X-ray Emission}",
      journal = {arXiv e-prints},
         year = 2026,
        month = jun,
          eid = {arXiv:2606.08931},
        pages = {arXiv:2606.08931},
archivePrefix = {arXiv},
       eprint = {2606.08931},
 primaryClass = {astro-ph.HE},
       adsurl = {https://ui.adsabs.harvard.edu/abs/2026arXiv260608931S}
}

@ARTICLE{2017ApJ...850..139W,
       author = {{Wong}, Tony and {Hughes}, Annie and {Tokuda}, Kazuki and {Indebetouw}, R{\'e}my and {Bernard}, Jean-Philippe and {Onishi}, Toshikazu and {Wojciechowski}, Evan and {Bandurski}, Jeffrey B. and {Kawamura}, Akiko and {Roman-Duval}, Julia and {Cao}, Yixian and {Chen}, C.-H. Rosie and {Chu}, You-hua and {Cui}, Chaoyue and {Fukui}, Yasuo and {Montier}, Ludovic and {Muller}, Erik and {Ott}, Juergen and {Paradis}, Deborah and {Pineda}, Jorge L. and {Rosolowsky}, Erik and {Sewi{\l}o}, Marta},
        title = "{ALMA Observations of a Quiescent Molecular Cloud in the Large Magellanic Cloud}",
      journal = {\apj},
         year = 2017,
        month = dec,
       volume = {850},
       number = {2},
          eid = {139},
        pages = {139},
          doi = {10.3847/1538-4357/aa9333},
archivePrefix = {arXiv},
       eprint = {1708.08890},
 primaryClass = {astro-ph.GA},
       adsurl = {https://ui.adsabs.harvard.edu/abs/2017ApJ...850..139W}
}

@ARTICLE{2011ApJS..197...16W,
       author = {{Wong}, Tony and {Hughes}, Annie and {Ott}, J{\"u}rgen and {Muller}, Erik and {Pineda}, Jorge L. and {Bernard}, Jean-Philippe and {Chu}, You-Hua and {Fukui}, Yasuo and {Gruendl}, Robert A. and {Henkel}, Christian and {Kawamura}, Akiko and {Klein}, Ulrich and {Looney}, Leslie W. and {Maddison}, Sarah and {Mizuno}, Yoji and {Paradis}, Deborah and {Seale}, Jonathan and {Welty}, Daniel E.},
        title = "{The Magellanic Mopra Assessment (MAGMA). I. The Molecular Cloud Population of the Large Magellanic Cloud}",
      journal = {\apjs},
         year = 2011,
        month = dec,
       volume = {197},
       number = {2},
          eid = {16},
        pages = {16},
          doi = {10.1088/0067-0049/197/2/16},
archivePrefix = {arXiv},
       eprint = {1108.5715},
 primaryClass = {astro-ph.GA},
       adsurl = {https://ui.adsabs.harvard.edu/abs/2011ApJS..197...16W}
}

@ARTICLE{2010PASJ...62...51M,
       author = {{Mizuno}, Yoji and {Kawamura}, Akiko and {Onishi}, Toshikazu and {Minamidani}, Tetsuhiro and {Muller}, Erik and {Yamamoto}, Hiroaki and {Hayakawa}, Takahiro and {Mizuno}, Norikazu and {Mizuno}, Akira and {Stutzki}, J{\"u}rgen and {Pineda}, Jorge L. and {Klein}, Uli and {Bertoldi}, Frank and {Koo}, Bon-Chul and {Rubio}, Monica and {Burton}, Michael and {Benz}, Arnold and {Ezawa}, Hajime and {Yamaguchi}, Nobuyuki and {Kohno}, Kotaro and {Hasegawa}, Tetsuo and {Tatematsu}, Ken'ichi and {Ikeda}, Masafumi and {Ott}, J{\"u}rgen and {Wong}, Tony and {Hughes}, Annie and {Meixner}, Margaret and {Indebetouw}, Remy and {Gordon}, Karl D. and {Whitney}, Barbara and {Bernard}, Jean-Philippe and {Fukui}, Yasuo},
        title = "{Warm and Dense Molecular Gas in the N 159 Region: $^{12}$CO J = 4-3 and $^{13}$CO J = 3-2 Observations with NANTEN2 and ASTE}",
      journal = {\pasj},
         year = 2010,
        month = feb,
       volume = {62},
       number = {1},
        pages = {51-67},
          doi = {10.1093/pasj/62.1.51},
archivePrefix = {arXiv},
       eprint = {0910.0309},
 primaryClass = {astro-ph.CO},
       adsurl = {https://ui.adsabs.harvard.edu/abs/2010PASJ...62...51M}
}

@ARTICLE{2006A&A...450..585B,
       author = {{Bamba}, A. and {Ueno}, M. and {Nakajima}, H. and {Mori}, K. and {Koyama}, K.},
        title = "{A detailed observation of a LMC supernova remnant DEM L241 with XMM-Newton}",
      journal = {\aap},
         year = 2006,
        month = may,
       volume = {450},
       number = {2},
        pages = {585-591},
          doi = {10.1051/0004-6361:20054096},
archivePrefix = {arXiv},
       eprint = {astro-ph/0601065},
 primaryClass = {astro-ph},
       adsurl = {https://ui.adsabs.harvard.edu/abs/2006A&A...450..585B}
}

@INPROCEEDINGS{2017ICRC...35..730K,
       author = {{Komin}, N. and {Haupt}, M. and {H.~E.~S.~S. Collaboration Affiliations: AA(University of the Witwatersrand}, AB(DESY Zeuthen), Johannesburg nukri. komin@wits. ac. za)},
        title = "{Discovery of VHE Gamma-Ray Emission from the Binary System LMC P3}",
    booktitle = {35th International Cosmic Ray Conference (ICRC2017)},
         year = 2017,
       series = {International Cosmic Ray Conference},
       volume = {301},
        month = jul,
          eid = {730},
        pages = {730},
          doi = {10.22323/1.301.0730},
archivePrefix = {arXiv},
       eprint = {1708.03171},
 primaryClass = {astro-ph.HE},
       adsurl = {https://ui.adsabs.harvard.edu/abs/2017ICRC...35..730K}
}

@ARTICLE{2012ApJ...759..123S,
       author = {{Seward}, F.~D. and {Charles}, P.~A. and {Foster}, D.~L. and {Dickel}, J.~R. and {Romero}, P.~S. and {Edwards}, Z.~I. and {Perry}, M. and {Williams}, R.~M.},
        title = "{DEM L241, a Supernova Remnant Containing a High-mass X-Ray Binary}",
      journal = {\apj},
         year = 2012,
        month = nov,
       volume = {759},
       number = {2},
          eid = {123},
        pages = {123},
          doi = {10.1088/0004-637X/759/2/123},
archivePrefix = {arXiv},
       eprint = {1208.1453},
 primaryClass = {astro-ph.HE},
       adsurl = {https://ui.adsabs.harvard.edu/abs/2012ApJ...759..123S}
}

@ARTICLE{1985ApJS...58..197M,
       author = {{Mathewson}, D.~S. and {Ford}, V.~L. and {Tuohy}, I.~R. and {Mills}, B.~Y. and {Turtle}, A.~J. and {Helfand}, D.~J.},
        title = "{Supernova remnants in the Magellanic Clouds. III.}",
      journal = {\apjs},
         year = 1985,
        month = jun,
       volume = {58},
        pages = {197-200},
          doi = {10.1086/191037},
       adsurl = {https://ui.adsabs.harvard.edu/abs/1985ApJS...58..197M}
}

@ARTICLE{2016ApJ...829..105C,
       author = {{Corbet}, R.~H.~D. and {Chomiuk}, L. and {Coe}, M.~J. and {Coley}, J.~B. and {Dubus}, G. and {Edwards}, P.~G. and {Martin}, P. and {McBride}, V.~A. and {Stevens}, J. and {Strader}, J. and {Townsend}, L.~J. and {Udalski}, A.},
        title = "{A Luminous Gamma-ray Binary in the Large Magellanic Cloud}",
      journal = {\apj},
         year = 2016,
        month = oct,
       volume = {829},
       number = {2},
          eid = {105},
        pages = {105},
          doi = {10.3847/0004-637X/829/2/105},
archivePrefix = {arXiv},
       eprint = {1608.06647},
 primaryClass = {astro-ph.HE},
       adsurl = {https://ui.adsabs.harvard.edu/abs/2016ApJ...829..105C}
}

@ARTICLE{2012ApJ...759...35I,
       author = {{Inoue}, Tsuyoshi and {Inutsuka}, Shu-ichiro},
        title = "{Formation of Turbulent and Magnetized Molecular Clouds via Accretion Flows of H I Clouds}",
      journal = {\apj},
         year = 2012,
        month = nov,
       volume = {759},
       number = {1},
          eid = {35},
        pages = {35},
          doi = {10.1088/0004-637X/759/1/35},
archivePrefix = {arXiv},
       eprint = {1205.6217},
 primaryClass = {astro-ph.GA},
       adsurl = {https://ui.adsabs.harvard.edu/abs/2012ApJ...759...35I}
}

@ARTICLE{2000ApJ...532..980K,
       author = {{Koyama}, Hiroshi and {Inutsuka}, Shu-Ichiro},
        title = "{Molecular Cloud Formation in Shock-compressed Layers}",
      journal = {\apj},
         year = 2000,
        month = apr,
       volume = {532},
       number = {2},
        pages = {980-993},
          doi = {10.1086/308594},
archivePrefix = {arXiv},
       eprint = {astro-ph/9912509},
 primaryClass = {astro-ph},
       adsurl = {https://ui.adsabs.harvard.edu/abs/2000ApJ...532..980K}
}

@ARTICLE{2019ApJ...873...40S,
       author = {{Sano}, H. and {Matsumura}, H. and {Nagaya}, T. and {Yamane}, Y. and {Alsaberi}, R.~Z.~E. and {Filipovi{\'c}}, M.~D. and {Tachihara}, K. and {Fujii}, K. and {Tokuda}, K. and {Tsuge}, K. and {Yoshiike}, S. and {Onishi}, T. and {Kawamura}, A. and {Minamidani}, T. and {Mizuno}, N. and {Yamamoto}, H. and {Inutsuka}, S. and {Inoue}, T. and {Maxted}, N. and {Rowell}, G. and {Sasaki}, M. and {Fukui}, Y.},
        title = "{ALMA CO Observations of Supernova Remnant N63A in the Large Magellanic Cloud: Discovery of Dense Molecular Clouds Embedded within Shock-ionized and Photoionized Nebulae}",
      journal = {\apj},
         year = 2019,
        month = mar,
       volume = {873},
       number = {1},
          eid = {40},
        pages = {40},
          doi = {10.3847/1538-4357/ab02fd},
archivePrefix = {arXiv},
       eprint = {1809.02481},
 primaryClass = {astro-ph.GA},
       adsurl = {https://ui.adsabs.harvard.edu/abs/2019ApJ...873...40S}
}

@ARTICLE{2018ApJ...867....7S,
       author = {{Sano}, H. and {Yamane}, Y. and {Tokuda}, K. and {Fujii}, K. and {Tsuge}, K. and {Nagaya}, T. and {Yoshiike}, S. and {Filipovi{\'c}}, M.~D. and {Alsaberi}, R.~Z.~E. and {Barnes}, L. and {Onishi}, T. and {Kawamura}, A. and {Minamidani}, T. and {Mizuno}, N. and {Yamamoto}, H. and {Tachihara}, K. and {Maxted}, N. and {Voisin}, F. and {Rowell}, G. and {Yamaguchi}, H. and {Fukui}, Y.},
        title = "{Molecular Clouds Associated with the Type Ia SNR N103B in the Large Magellanic Cloud}",
      journal = {\apj},
         year = 2018,
        month = nov,
       volume = {867},
       number = {1},
          eid = {7},
        pages = {7},
          doi = {10.3847/1538-4357/aae07c},
archivePrefix = {arXiv},
       eprint = {1806.10299},
 primaryClass = {astro-ph.GA},
       adsurl = {https://ui.adsabs.harvard.edu/abs/2018ApJ...867....7S}
}

@ARTICLE{2014ApJ...781...70F,
       author = {{Furukawa}, N. and {Ohama}, A. and {Fukuda}, T. and {Torii}, K. and {Hayakawa}, T. and {Sano}, H. and {Okuda}, T. and {Yamamoto}, H. and {Moribe}, N. and {Mizuno}, A. and {Maezawa}, H. and {Onishi}, T. and {Kawamura}, A. and {Mizuno}, N. and {Dawson}, J.~R. and {Dame}, T.~M. and {Yonekura}, Y. and {Aharonian}, F. and {de O{\~n}a Wilhelmi}, E. and {Rowell}, G.~P. and {Matsumoto}, R. and {Asahina}, Y. and {Fukui}, Y.},
        title = "{The Jet and Arc Molecular Clouds toward Westerlund 2, RCW 49, and HESS J1023-575 $^{12}$CO and $^{13}$CO (J = 2-1 and J = 1-0) observations with NANTEN2 and Mopra Telescope}",
      journal = {\apj},
         year = 2014,
        month = feb,
       volume = {781},
       number = {2},
          eid = {70},
        pages = {70},
          doi = {10.1088/0004-637X/781/2/70},
archivePrefix = {arXiv},
       eprint = {1401.4845},
 primaryClass = {astro-ph.GA},
       adsurl = {https://ui.adsabs.harvard.edu/abs/2014ApJ...781...70F}
}
\bibliographystyle{aasjournalv7}

%% This command is needed to show the entire author+affiliation list when
%% the collaboration and author truncation commands are used.  It has to
%% go at the end of the manuscript.
%\allauthors

%% Include this line if you are using the \added, \replaced, \deleted
%% commands to see a summary list of all changes at the end of the article.
%\listofchanges

\end{document}